# Spin splitting torque enabled artificial neuron with self-reset via synthetic antiferromagnetic coupling

Badsha Sekh[1], Hasibur Rahaman[1], Ravi Shankar Verma[2], Ramu Maddu[1], Kesavan Jawahar[1] and S.N. Piramanayagam[1,*]

[1]School of Physical and Mathematical Sciences, Nanyang Technological University, 21 Nanyang Link, 637371, Singapore

[2]Indian Institute of Technology Roorkee, Roorkee, Uttarakhand, India – 247667

*Corresponding author: prem@ntu.edu.sg

Abstract

Spintronic artificial neurons are intriguing building blocks for energy efficient Neuromorphic Computing (NC). Nevertheless, most contemporary implementations rely on symmetry breaking external in plane magnetic fields ($H_X$) for neuron operation, which limits scalability and hardware practicality. We experimentally demonstrate an altermagnet/Synthetic Antiferromagnetic Coupling (SAF) based spintronic neuron that uses out of plane spin ($\sigma_Z$) polarized spin-splitting torque to eliminate the necessity of an external $H_X$. The neuron device also features intrinsic self-reset function facilitated by built-in exchange coupling. Furthermore, the proposed device is validated for Spiking Neural Network (SNN) applications by achieving test accuracies of 95.99% and 94.36% on the MNIST and N-MNIST datasets, respectively. These results demonstrate the hardware feasibility and compatibility of the proposed spintronic neuron, highlighting its potential for compact, scalable and energy-efficient neuromorphic computing systems.

Despite the rapid growth of Artificial Intelligence (AI) over the past decade, further improvements in computing efficiency are increasingly constrained due to slowdown in the device scaling and high cost of big data movements, known as the von Neumann bottleneck[1]. Brain-inspired computing has emerged as a new computing paradigm, offering key advantages such as low power consumption, high-speed operation, and massive parallelism. In recent years, fundamental elements of human brain, such as neurons and synapses have been emulated utilizing resistive switching memory[2,3], phase-change memory[4,5], ferroelectric memory[6,7], and spintronic devices[8,9] as potential building blocks for NC hardware. In particular, spintronic devices hold a promise for hardware implementation owing to their non-volatility, high-speed operation, and energy-efficient domain wall motion employing spin orbit torque (SOT) effect. Although, significant progress has been made in emulating synaptic functionalities, realizing power-efficient leaky integrate-and-fire capabilities in artificial neurons remains a challenge. To address this issue, artificial neurons[10,11] based on magnetic skyrmions[12,13] have recently been demonstrated through micromagnetic simulations. However, practical limitations including room temperature stability, the skyrmion Hall effect[14], and skyrmionic stochastic dynamics[15] hinder its use for large-scale NC. Alternatively, different material platforms including nano oscillator[16,17], domain wall based ferromagnets[18,19] etc have been explored. Nevertheless, most existing neuron designs rely on external magnetic fields or additional circuitry to facilitate reset operations, which increases system complexity and power consumption.

In this context, SAFs [20–23] with strong antiferromagnetic interlayer coupling offer a possible solution. The typical SAF stack consists of a bottom ferromagnetic (FM) layer which acts as soft layer and a top FM layer serving as hard layer, coupled antiferromagnetically through an intrinsic interlayer exchange field ($H_{ex}$). Therefore, $H_{ex}$ drives the magnetization of the free layer back to its resting state after firing, emulating the reset of an action potential in biological neurons. However, conventional SAF based neurons[24–26] typically require large external assistance magnetic fields to reliably induce magnetization switching during integration process, which limits their practical scalability for large neuromorphic arrays.

Collectively, these studies highlight the necessity of a neuron design that is compact, field-free, energy-efficient, and capable of self-reset, without sacrificing the scalability. In this regard, altermagnets[27] have attracted growing interest as a potential source of $\sigma_Z$ spin polarization, enabling deterministic magnetization switching[28,29] without any external $H_X$, overcoming a significant bottleneck in conventional spintronic neurons. In the search for suitable altermagnetic materials, $RuO_2$ has emerged as a promising candidate owing to its unique band structure. In $RuO_2$, Ru atoms at two magnetic sublattices are surrounded by O atoms with distinct local environments, resulting in symmetry breaking[30] that gives rise to spin-split electronic bands. When we apply a charge current in a certain direction through altermagnet, the shifting of Fermi surface may lead to generation of an odd time-reversal transverse spin current due to spin splitting effect (SSE). Unlike the Spin Hall Effect[31] (SHE), the spin polarization in this case follows the direction of the Néel vector. Since, $RuO_2$ (101) possesses a Néel vector canted towards the z direction, the resulting spin splitting induced torque with tilted $\sigma_Z$ components[32] will break the magnetic domain wall symmetry enabling field free deterministic magnetization switching.

In this work, we leverage the spin splitting properties of altermagnets for the first time to demonstrate a SAF based spintronic neuron without the need of any $H_X$ by integrating two spin sources, namely altermagnetic $RuO_2$ and Pt SOT layer. Upon application of in-plane charge current ($J_C$), the SOTs generated from two distinctive but cooperative mechanisms, facilitate magnetization switching at $H_X$ = 0 Oe. Firstly, the Pt layer generates $\sigma_Y$ spin polarization producing Damping Like (DL) torque due to conventional SHE. Second, a Crytal angle dependent $\sigma_Z$ polarized spin splitting torque emerges in $RuO_2$. During the application of the current pulses, the combined torques from $RuO_2$ and Pt compete against the $H_{ex}$ of the SAF. Once the accumulated torques exceed the $H_{ex}$ and switching field, the soft layer (reddish purple colour spins) undergoes deterministic magnetization reversal as schematically shown in Figure 1a. This magnetization reversal corresponds to the integration activity of the neuron. When we turn off the input current, the soft layer restores back to its initial magnetic configuration due to $H_{ex}$, mimicking the reset functionality.

In the present study, to experimentally realize the aforementioned phenomena, we first sputter $RuO_2$ (101) thin film in presence of a mixed gas environment (12 sccm Ar and 8 sccm $O_2$) on single crystalline $Al_2O_3$ (1-102) substrate. The Θ-2Θ XRD data (Figure 1b) confirms the epitaxial growth of $RuO_2$ thin film.

**Table1:** Full SAF stack details having different thin layers (layer thicknesses are in nm).

| Samples | Deposited stack |
|---|---|
| S1 | $Al_2O_3$/$RuO_2$(18)/Pt(1.4)/Co(0.5)/Pt(0.5)/Co(0.5)/Ru(2.5)/[Co(0.5)/Pt(0.8)]$_{×2}$/Ru(2) |
| S2 | $Al_2O_3$/$RuO_2$(18)/Pt(1.3)/Co(0.5)/Pt(0.8)/Co(0.5)/Ru(2.6)/[Co(0.5)/Pt(0.8)]$_{×2}$/Ru(2) |
| S3 | $Al_2O_3$/$RuO_2$(18)/Pt(1.3)/Co(0.4)/Ni(0.95)/Co(0.4)/Ru(2.30)/[Co(0.4)/Pt(0.8)]$_{×4}$/Ru(2) |
| S4 | $Al_2O_3$/$RuO_2$(18)/Pt(1.3)/Co(0.4)/Ni(1)/Co(0.4)/Ru(2.35)/[Co(0.4)/Pt(0.8)]$_{×4}$/Ru(2) |
| S5 | $Al_2O_3$/$RuO_2$(18)/Pt(1.4)/Co(0.4)/Ni(1)/Co(0.4)/Ru(2.35)/[Co(0.4)/Pt(0.8)]$_{×4}$/Ru(2) |

Furthermore, we have deposited a series of SAF stacks (Table 1) to obtain an optimal $H_{ex}$ as well as a lower magnetization switching field comprising Pt as an additional SOT source on the epitaxially grown $RuO_2$ (101) layer. Here, the bottom layer of Ru acts as a soft layer whereas the upper layer acts as a hard layer. A representative Magneto Optical Kerr Effect (MOKE) hysteresis loops of thin film sample (S5) is shown in Figure 1c, demonstrating sharp perpendicular magnetic anisotropy (PMA). MOKE measurements for all other samples, along with vibrating sample magnetometry (VSM) results are provided in the supplementary information S1, confirm robust PMA across the entire sample series.

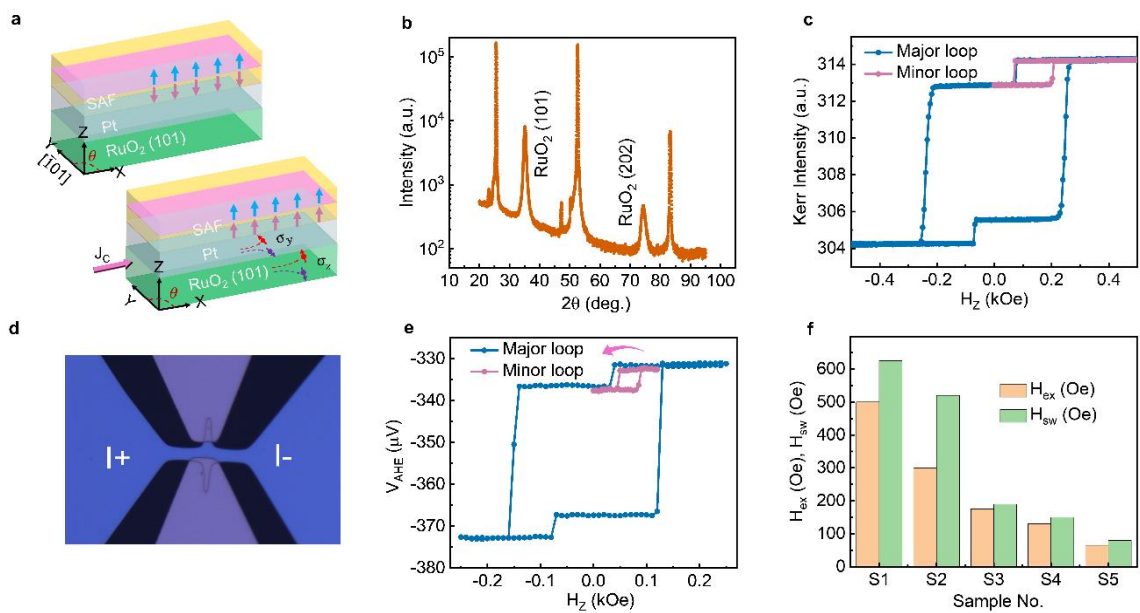

**Figure 1.** Device concept, structural characterization, and magnetic properties of the SAF spintronic neuron. (a) Schematic of the proposed mechanism showing soft layer magnetization switching. (b) Θ-2Θ XRD data of the deposited $RuO_2$ (18 nm) thin film on $Al_2O_3$ (1-102) substrate.

(c) Kerr hysteresis loop of the SAF stack having sharp PMA. (d) Optical image of the fabricated Hall bar device. (e) AHE loops of the patterned Hall bar device. (f) Exchange field ($H_{ex}$) and switching field ($H_{sw}$) across various samples.

Subsequently, the Hall bar devices (Figure 1d) are fabricated comprising the SAF structure using optical lithography followed by ion milling. The switching field ($H_Z^{SW}$) of the soft layer across different $H_{ex}$ has been measured (as shown in Figure 1f) from the Anomalous Hall effect (AHE) minor loop measurements (Figure 1e). To isolate and validate the role of the SOT in magnetization switching prior to neuron operation, we next investigate current-induced magnetization dynamics in Hall-bar devices comprising only the soft ferromagnetic layer. This approach allows us to examine switching mechanism independent of interlayer exchange coupling. The Kerr hysteresis loop of the soft layer having a sharp PMA is shown in the supplementary information S2. As displayed in Figure 2a, the squareness of the AHE loop indicates the PMA of the studied system. To probe the current-driven response, the device has been initially saturated in a positive magnetization state and allowed to relax to its remanent state. Motivated by our previous experimental findings[33], which reveal a maximized spin-splitting torque when the charge current is applied along the 90° crystallographic direction with respect to RuO$_2$ [$\bar{1}$01], we sweep pulsed DC current (I) (as schematically shown in Figure 2b) at this specific orientation without any applied $H_X$. The write and read current pulse widths are fixed at 300 µs and 0.15 s, respectively. The change in AHE voltage ($V_{AHE}$) increases monotonically with increasing current and a gradual 100% field free magnetization switching is achieved in correspondence to maximum applied current. The switching probability is defined as $\frac{\Delta V_H}{\Delta V_H^{max.}}$, where $\Delta V_H$ is the change in $V_{AHE}$ induced by successive current pulses and $\Delta V_H^{max}$ indicates the $V_{AHE}$ difference between two opposite magnetization states, as determined from AHE measurement.

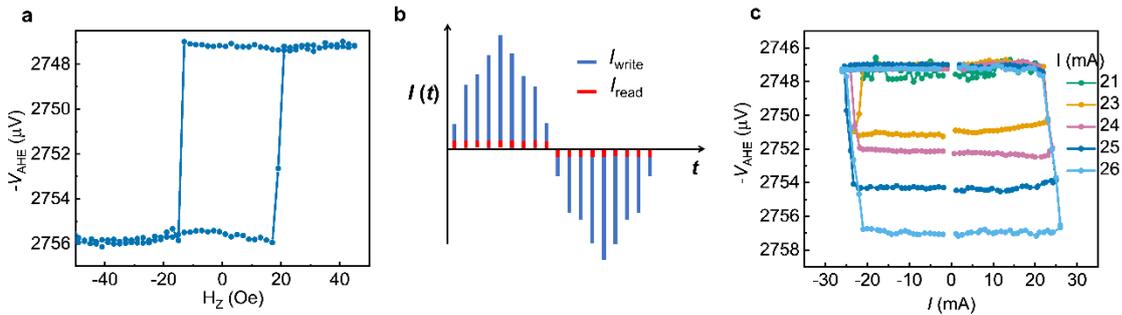

**Figure 2.** Electrical characterization and current-driven switching behavior of the soft layer. (a) AHE loop of the Hall bar device comprising only soft layer. (b) Schematic measurement protocol during current pulse application for magnetization switching. (c) Field free magnetization switching at different applied currents.

Having established the $\sigma_Z$ induced SOT field is sufficient to break the down-up and up-down domain wall symmetry and enable field free switching of the soft magnetic layer, we subsequently extend neuron functionality incorporating the complete SAF stack following same measurement protocols. In the remanent magnetization state of the full stack, the magnetizations of the soft and hard layers are aligned antiparallel. Therefore, to achieve full integration functionality of the neuron hall bar device, we need soft layer magnetization switching driving parallel configuration. In contrast to the only soft layer switching dynamics, complete magnetization reversal of the soft layer in full SAF stack requires a small out of plane assistance field ($H_Z^{assist.}$) to overcome the $H_{ex}$ (switching measurements with different $H_Z^{assist.}$ is shown in supplementary information S3), as shown schematically in Figure 3a. Notably, the $H_Z^{assist.}$ remains always lower than $H_{sw}$ of the soft layer magnetization, ensuring SOT induced manipulation of its magnetization. The trend of need of $H_Z^{assist.}$ increases with increasing $H_Z^{SW}$ for various $H_{ex}$ as it has been exhibited in Figure 3b. The integration function owing to the smallest $H_Z^{assist.}$ (50 Oe) correspondence to $H_{ex}$ = 65 Oe has been shown in Figure 3c. The integration function is quantified as switching percentage, which increases with increasing current density due to SOT field arising from spin splitting effect in $RuO_2$ layer and DL field generated in Pt layer. Here, the $\sigma_Y$ contribution from $RuO_2$ layer is almost negligible[34] due to its lower spin hall angle compared to Pt. The corresponding MOKE images at different magnetization switching stages have been shown in Figure 3d. Here, note of that we do not apply $H_X$ during the measurement. It indicates that the spin splitting induced torque is sufficient to break the symmetry, thereby enabling magnetization switching. Additionally, we investigate the self-reset property and impact of applied $H_X$ during integration while maintaining $H_Z^{assist.}$ = 50 Oe. The magnetization reversal happens even at lower current density (J) due to applied $H_X$ parallel to the current. The trend of required J for magnetization reversal reduces monotonically with increasing the magnitude of the $H_X$ (Figure 3e). These results are ascribed to the additional symmetry breaking field contribution from applied $H_X$, which enables magnetization switching faster. Crucially, we also demonstrate that the self-reset function due to $H_{ex}$ has been achieved in all cases following the integration activity. Moreover, this reset process is entirely intrinsic to the device and does not rely on any auxiliary reset pulses or external circuit components.

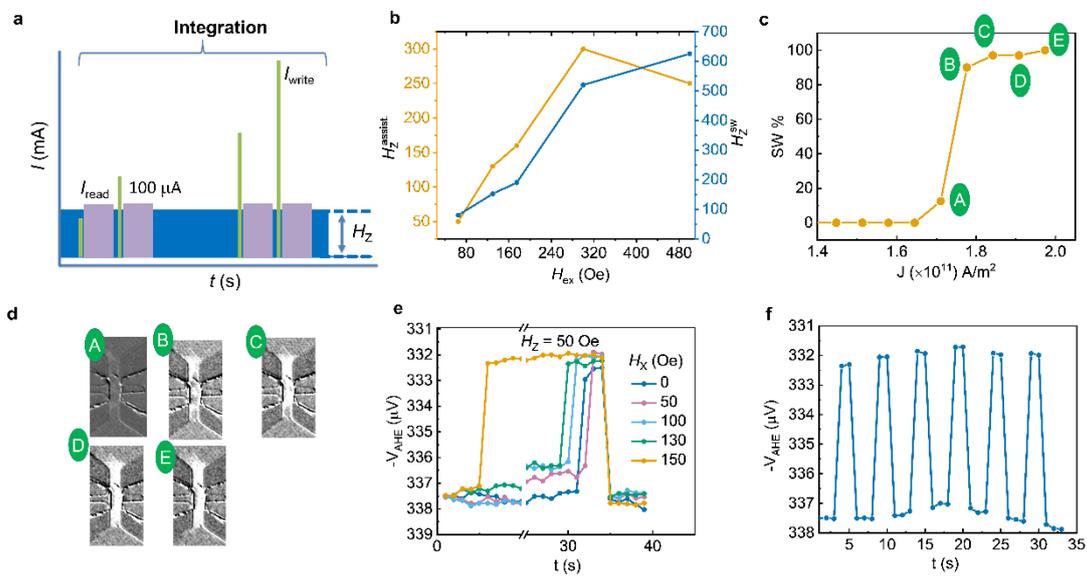

**Figure 3.** Measurement scheme and operational dynamics of the neuron device. (a) Schematic of the measurement protocol for neuron operation. (b) Variation of $H_Z^{assist.}$ correspondence to various $H_{ex}$. (c) SOT driven integration function under different applied current density (J). (d) Kerr images during magnetization switching of the soft layer under applied successive currents. (e)

Self-reset property and integration function as a function of $H_X$. (f) Reproducibility of the neuron operation in several cycles.

We also show the reproducibility of neuron Hall bar device functionality in 6 repetition cycles (Figure 3f). Near the firing threshold condition, when the competition between the SOT and switching field becomes comparable, thermal fluctuations induce subtle stochastic magnetization dynamics over repeated cycles, resulting in small variations ($\frac{\sigma}{\mu} \times 100 \leq 0.05$ %) in the maximum switching probability. Here, σ and μ represent the standard deviation and average value of $V_{AHE}$ in repetition cycles, respectively. Almost the same integration and reset time in each cycle manifests excellent stability of the neuron dynamics.

We further elucidate the underlying physics of the proposed concept using micromagnetic simulations[35] to study $\sigma_z$-polarized spin-torque-driven DW dynamics in an altermagnet based SAF stack. Figure 4a shows the simulated DW device consisting of two perpendicularly magnetized ferromagnetic layers, FM1 (soft) and FM2 (hard) with lateral dimensions of 320 nm × 32 nm, which are coupled via an effective SAF interaction implemented through region-dependent exchange scaling. A DW is intentionally introduced at 32 nm distance from the left end by initializing a reverse domain. DW motion occurs only in FM1 while FM2 remains fixed and provides an SAF-induced $H_{ex}$, enabling controlled and reversible reset behavior.

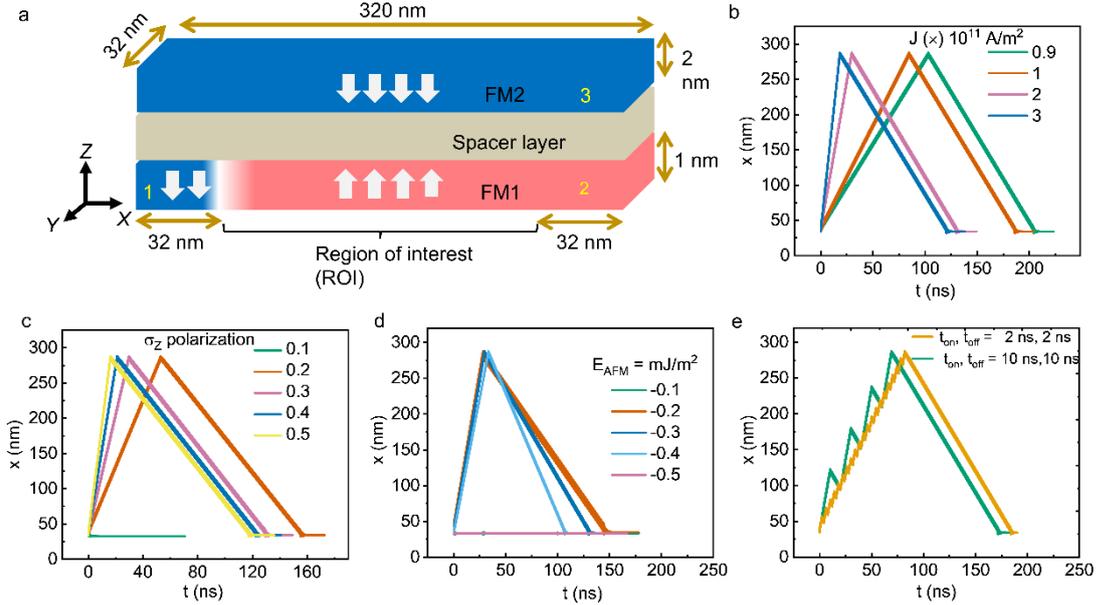

**Figure 4.** Micromagnetic simulations of domain-wall (DW) dynamics in a synthetic antiferromagnetic heterostructure due to $\sigma_z$-polarized spin-torque. (a) Schematic of the simulated DW device with a soft layer (FM1) and a hard layer (FM2) coupled via SAF interaction; DW motion occurs only in FM1. (b) DW position (X) versus time (t) for different current densities (J). J is varied from 0.9×10¹¹ A/m² to 3×10¹¹ A/m². (c) Effect of $\sigma_z$ polarization strength on DW dynamics. (d) Influence of SAF coupling strength ($E_{AFM}$) on DW integration and reset. (e) DW dynamics for different current pulse durations, showing stepwise integration with leakage followed by robust reset behaviour.

In the above mentioned SAF framework (with the interlayer exchange energy density of $E_{\text{AFM}}$ = -0.3 mJ/m²), an applied $J_C$ generates a σ$_z$-polarized DL spin torque on the magnetic moment (m) described by m × (m × $\sigma_z$), which drives DW motion by favoring one particular domain over the other[36,37]. Figure 4b shows the DW position (X) as a function of time (t) for current densities (J) ranging from $0.9 \times 10^{11}$ to $3 \times 10^{11}$ A m$^{-2}$. Higher J produces steeper slopes in t versus X graph, indicating faster DW propagation. This behavior is attributed to the enhanced σ$_z$ induced spin torque, which more effectively overcomes the $H_{ex}$ opposing force and intrinsic pinning potential. The continuous DW displacement under applied J emulates the integration process, while the maximum displacement is limited to 288 nm to avoid any DW annihilation. Subsequently after removing J, the DW relaxes back to its initial nucleated region due to the SAF-induced restoring torque, demonstrating robust and reversible reset behavior following each integration processes.

Next, we investigate the effect of various σ$_z$ polarization strengths on DW dynamics during integration process for a particular J = 2 × 10$^{11}$ A/m² (Figure 4c). For a σ$_z$ polarization value corresponding to a spin Hall angle of 0.1, no DW motion is observed, indicating the applied torque is insufficient at this value to overcome the intrinsic pinning potential[38]. As higher σ$_z$ polarization leads to faster DW motion, an appreciable DW displacement occurs only beyond its certain threshold value. However, the DW exhibits a clear reset operation after all σ$_z$ induced integration process. Notably, the reset velocities in Figures 4b and 4c remain nearly constant across all the events. In addition, the extracted integration and reset velocities are summarized in supplementary information S4.

Building on these results, we analyze the DW dynamics under various $E_{\text{AFM}}$ (as shown in Figure 4d) using a fixed J = 2 × 10$^{11}$ A/m² and spin Hall angle = 0.2 for σ$_z$ component. The current driven forward DW displacement is affected by $E_{\text{AFM}}$, as the SAF-induced torque opposes the σ$_z$ polarized DL torque driving the DW. Therefore, an optimized coupling strength of $E_{\text{AFM}}$ = -0.3 mJ/m² has been selected in most of the simulations while varying other parameters. For weaker SAF coupling ($E_{\text{AFM}}$ = -0.2 mJ/m²), DW motion becomes easier, whereas increasing the coupling strength ($E_{\text{AFM}}$ = -0.4 mJ/m²) impedes the DW propagation. At a critical value of $E_{\text{AFM}}$ = -0.5 mJ/m², the restoring force dominates and DW motion is effectively suppressed. In contrast, the reset DW dynamics exhibits the opposite relation with respect to $E_{\text{AFM}}$ described as: higher | $E_{\text{AFM}}$ | accelerates DW to return to its initial position after removal of current owing to a stronger SAF-induced restoring torque, whereas weaker coupling leads to a slower reset operation. These results clearly show that the SAF coupling controls the reset process and can be tuned independently without changing the driving *J*.

We now compare DW dynamics for two pulse schemes, $t_{\text{on}} = t_{\text{off}} = 2$ ns and $t_{\text{on}} = t_{\text{off}} = 10$ ns as demonstrated in Figure 4e. In both cases, the DW advances in a stepwise manner, with discrete forward displacement during each pulse and partial relaxation (leakage) during the off-time due to SAF-induced restoring torque. Longer pulses yield smoother DW motion by providing a more sustained driving torque and reduced leakage. Despite these differences, both pulse schemes achieve comparable forward displacement, and the DW reliably resets to its initial pinned position after current removal, governed by the SAF coupling. These results manifest that the σ$_z$ induced DW integration dynamics can be tailored via the pulse parameters, while the reset behavior remains robust and is primarily governed by intrinsic $E_{\text{AFM}}$.

In order to assess the adaptability of the proposed neuron device to NC applications, a SNN on the adaptive leaky integrate-and-fire (LIF) neuron dynamics is introduced to classify handwritten digits in event-based applications. Unlike in the classical LIF model, the neuron model that is employed in this work is subjected to two independent time constants that are the integration and leakage time constants. This dissociation enables independent tuning of the integration and reset processes, providing greater flexibility to engineer device-level dynamics and to tailor neuronal responses to temporally sparse, event-driven inputs[39].

The N-MNIST dataset consists of 70,000 samples, including 60,000 training samples and 10,000 test samples, distributed across ten-digit classes (0–9). It is an event-based variant of the standard MNIST dataset, introduced by Orchard et al.[40] and proposing a set of fixed MNIST images on a screen and having a Dynamic Vision Sensor (DVS) execute predefined saccadic movements. These motions cause relative motion between the sensor and the image which forms asynchronous ON and OFF events that code pixel intensity variation with time. This event-driven representation is used to train and evaluate the network and maintain the time-related information and resembles realistic neuromorphic vision sensing[40].

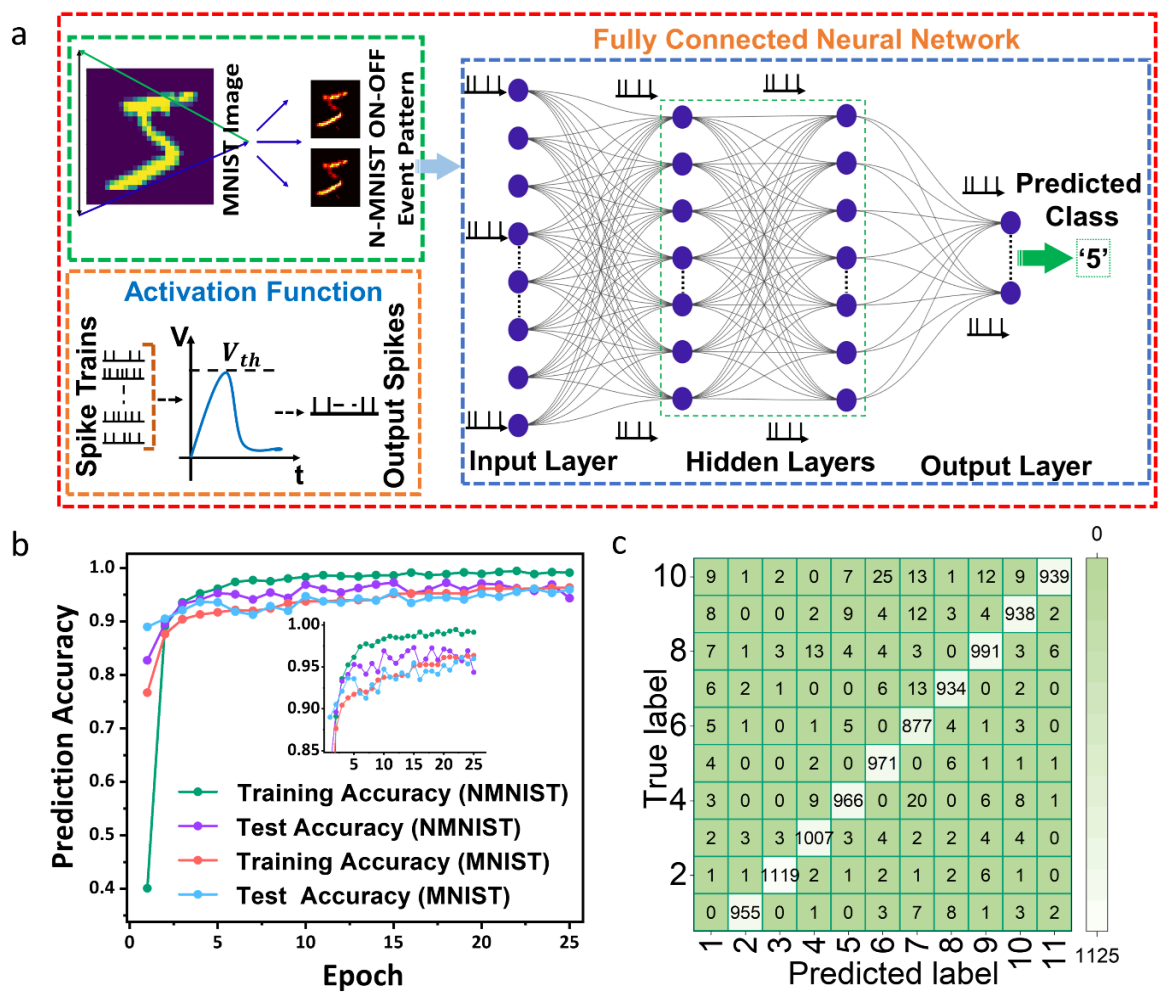

**Figure 5**. Schematic illustration of the neural network architecture used to explore SNNs for handwritten digit classification on the NMNIST dataset. (a) Each original 28 × 28 pixel MNIST image with ON and OFF polarity events, is first converted into discrete temporal frames. These temporal frames are then encoded sequentially into spike trains, fed to fully connected neuron network of LIF neuron trained using the surrogate gradient backpropagation (BP) (b) Training and testing accuracy curves for the classification of N-MNIST and MNIST datasets (c) Confusion matrix for N-MNIST, illustrating class-wise prediction accuracy and misclassification trends.

The network architecture employed for N-MNIST classification has been illustrated in Figure 5a. The input layer is the first layer consisting of neurons corresponding to the spatial (x, y) resolution of the stabilized N-MNIST event stream. Each input neuron represents a single pixel location and accumulates incoming ON and OFF events within fixed temporal bins to form time-binned input representations, which are then processed sequentially by the SNN. Each input neuron is fully connected to the subsequent layer of spiking neurons via synaptic weights. These hidden spiking layers act as feature extractors, integrating time-binned event information over time. During supervised training using surrogate-gradient–based BP, synaptic weights are continuously updated, enabling individual neurons to become tuned to specific input patterns, such as characteristic spatial and temporal features of handwritten characters. The output layer consists of spiking neurons corresponding to the target digit classes. During inference, the spiking activity of each output neuron is accumulated over the full observation window, and the final classification decision is obtained by selecting the neuron with the highest cumulative spike count. In our adaptive LIF Neuron Model, the membrane potential (V(t)) of the neuron k at time t in the full connected network layer can be described as

$$\frac{dV_k(t)}{dt} = -\frac{(V_k(t) - V_{reset})}{\tau_{leak}} + \frac{-V_k(t) + RI_k(t)}{\tau_{integration}} \quad (1)$$

where V(t) is the membrane potential, $I_k(t)$ is the total synaptic input current, R is the membrane resistance, and $\tau_{integration}$ and $\tau_{leak}$ denote the integration and leakage time constants, respectively. When the membrane potential surpasses the threshold ($V_{th}$), neuron emits a spike. Following spike generation, the membrane potential is reset to a fixed reset voltage to ensure stable spiking behavior. The Table 2 summarizes the key neuron parameters used in the implementation of the proposed SNN architecture. All simulation parameters were taken from experimental measurements, except for the time constants. Due to measurement limitations, millisecond time scales could not be accessed experimentally; however, such time scales are well established in the literature[41,42]. All neurons in the network share identical LIF membrane dynamics, with their functional roles determined solely by the learned synaptic weights in a feedforward architecture.. The Synaptic currents generated by presynaptic spikes are summed and directly integrated by the postsynaptic neuron according to the LIF dynamics[43].

Table 2: The parameters utilized in the implementation of the SNN architecture.

| Parameters | Threshold Voltage | Reset Voltage | Integration Time Constant | Leak Time constant |
|---|---|---|---|---|
| **Values** | 332.11 $\mu V$ | 336.52 $\mu V$ | 0.6492 ms | 0.8 ms |

During training and inference, N-MNIST event streams are discretized into millisecond-scale time bins, with events occurring within each bin accumulated to form temporal frames. These frames are processed sequentially by the spiking neural network over a fixed observation window, using the same pre-processing and network settings for consistency[44,45]. Spike generation is enabled via surrogate-gradient methods, allowing gradient-based training while retaining spike-based computation. Moreover, classification is based on the cumulative spike count of output neurons, with the predicted class given by the neuron showing the highest total spiking activity during inference.

Table 3: Classification accuracy of spiking neural networks on MNIST and N-MNIST datasets.

| Architecture | Dataset | Pre-Processing | Neuron Model | Training Type | Learning Rule | Prediction Accuracy (%) |
|---|---|---|---|---|---|---|
| Two-layer SNN[49] | MNIST | None | LIF (E–I, WTA) | Unsupervised | Exponential STDP | 95.0 |
| STDP SNN [47] | N-MNIST | Event stream | IF | Unsupervised | STDP + E–I | 80.6 |
| STDP Hardware SNN[50] | Iris | None | DW-based IF | Un/Part-Supervised | STDP + Homeostasis | 96.7 |
| DSNN[51] | MNIST | Kernel smoothing | LIF | Supervised | Spike-train BP | 95.65 |
| Deep-STDP SNN[52] | Tiny-ImageNet (10-class) | CNN features | LIF | Unsupervised (Hybrid) | STDP + Clustering | 56.97 |
| Proposed SNN (This work) | MNIST | None | Adaptive LIF (dual-$\tau$) | Supervised | Surrogate-gradient BP | 95.99 |
| Proposed SNN (This work) | N-MNIST | Time-binned frames | Adaptive LIF (dual-$\tau$) | Supervised | Surrogate-gradient BP | 94.36 |

**E-I: Excitatory-Inhibitory, IF: Integrate and Fire.

Although spike-timing-dependent plasticity (STDP) is widely used to study neuronal response behavior and feature learning, this work evaluates performance using a supervised SNN trained via surrogate-gradient BP. This learning strategy enables high classification accuracy while preserving event-driven inference and hardware-compatible LIF neuron dynamics[45]. Table 3 summarizes the classification performance of representative supervised and unsupervised SNN models reported in the literature across MNIST and N-MNIST benchmarks. Early SNN models employing unsupervised STDP with winner-take-all and excitatory–inhibitory competition typically reached 93–97% accuracy on MNIST. More recent supervised spike-based methods, such as spike-train BP, have reported comparable performance, while conversion-based approaches reach higher accuracy at the cost of reduced biological plausibility due to non-spiking training stages.

In contrast, our proposed SNN is trained directly in the spiking domain using supervised surrogate-gradient BP, preserving LIF neuron dynamics and spike-based inference. Using this framework, the network achieves test accuracy of 95.99% on MNIST, which is competitive with both supervised spike-based and unsupervised STDP-based methods despite its simple fully connected architecture. On the more challenging N-MNIST dataset, which exhibits temporal variability, sensor noise, and sparse event-driven inputs, the model attains a test accuracy of 94.36%, (Figure 4b). Furthermore, the confusion matrix in Figure 4c demonstrates the behaviour of class-wise prediction and misclassification patterns, that proves the ability of the model to predict digit classes reliably. The modest performance reduction compared to MNIST is consistent with prior neuromorphic studies. Overall, these findings indicate that the proposed adaptive LIF-based SNN offers an appropriate trade-off between classification accuracy and neuromorphic efficiency, making it well suited for event-driven vision devices and hardware-based implementations.

In summary, we report an altermagnet/SAF-based artificial neuron that operates without external $H_X$ and exhibits self-reset behavior enabled by $H_{ex}$. The experimentally observed integration and reset mechanisms are further validated through micromagnetic simulations. Moreover, the proposed spintronic neuron is evaluated within a SNN framework using the MNIST and N-MNIST datasets, achieving classification accuracies of 95.99% and 94.36%, respectively. These results demonstrate the compatibility of the device with SNN-based inference and learning. Taken together, this work establishes the proposed heterostructure as a viable and scalable platform for compact, energy-efficient, and $H_X$-free spintronic neurons, representing a significant step toward practical neuromorphic hardware integration.

1. **Acknowledgements**

The authors gratefully acknowledge the funding from the National Research Foundation (NRF), Singapore for CRP-frontier grant NRF-F-CRP-2024-0012. BS acknowledges the financial assistance from the NTU research scholarship.

2. **Author contributions**

Badsha Sekh: Sample preparation, device fabrication, design of experiments, device testing, analysis, interpretation and writing. Hasibur Rahaman: Device testing, design of experiments, analysis, micromagnetic simulations, and supported writing. Ravi Shankar Verma: Design and Implementation of a Fully Connected SNN for Image Classification and supported writing. Ramu Maddu: Device testing and analysis. Kesavan Jawahar: Device testing. S.N. Piramanayagam: Conceptualization of the main idea, design of experiments, analysis, interpretation, and writing.

3. **Conflict of Interest**

The authors declare no conflict of interest.

4. **Data Availability Statement**

The data that support the findings of this study are available from the corresponding author upon reasonable request.

**References:**

1. Syed, G. S., Le Gallo, M., & Sebastian, A. (2024). Non von neumann computing concepts. In Phase Change Materials-Based Photonic Computing (pp. 11-35). Elsevier.

2. Deshmukh, Sanchit, Miguel Muñoz Rojo, Eilam Yalon, Sam Vaziri, Cagil Koroglu, Raisul Islam, Ricardo A. Iglesias, Krishna Saraswat, and Eric Pop. 'Direct measurement of nanoscale filamentary hot spots in resistive memory devices.' Science Advances 8, no. 13 (2022): eabk1514.

3. Lin, Wen-Peng, Shu-Juan Liu, Tao Gong, Qiang Zhao, and Wei Huang. 'Polymer-based resistive memory materials and devices.' Advanced materials 26, no. 4 (2014): 570-606.

4. Lee, Se-Ho, Yeonwoong Jung, and Ritesh Agarwal. 'Highly scalable non-volatile and ultra-low-power phase-change nanowire memory.' Nature nanotechnology 2, no. 10 (2007): 626-630.

5. Park, See-On, Seokman Hong, Su-Jin Sung, Dawon Kim, Seokho Seo, Hakcheon Jeong, Taehoon Park, Won Joon Cho, Jeehwan Kim, and Shinhyun Choi. 'Phase-change memory via a phase-changeable self-confined nano-filament.' Nature (2024): 1-6.

6. Feng, Guangdi, Qiuxiang Zhu, Xuefeng Liu, Luqiu Chen, Xiaoming Zhao, Jianquan Liu, Shaobing Xiong et al. 'A ferroelectric fin diode for robust non-volatile memory.' Nature Communications 15, no. 1 (2024): 513.

7. Park, Ju Yong, Duk-Hyun Choe, Dong Hyun Lee, Geun Taek Yu, Kun Yang, Se Hyun Kim, Geun Hyeong Park et al. 'Revival of ferroelectric memories based on emerging fluorite-structured ferroelectrics.' Advanced Materials 35, no. 43 (2023): 2204904.

8. Kumar, Durgesh, Hong Jing Chung, JianPeng Chan, Tianli Jin, Sze Ter Lim, Stuart SP Parkin, Rachid Sbiaa, and S. N. Piramanayagam. 'Ultralow energy domain wall device for spin-based neuromorphic computing.' ACS nano 17, no. 7 (2023): 6261-6274.

9. Song, Kyung Mee, Jae-Seung Jeong, Biao Pan, Xichao Zhang, Jing Xia, Sunkyung Cha, Tae-Eon Park et al. 'Skyrmion-based artificial synapses for neuromorphic computing.' Nature Electronics 3, no. 3 (2020): 148-155.

10. Li, S., Kang, W., Huang, Y., Zhang, X., Zhou, Y., & Zhao, W. (2017). Magnetic skyrmion-based artificial neuron device. Nanotechnology, 28(31), 31LT01.

11. Lone, A. H., Rahimi, D. N., Fariborzi, H., & Setti, G. (2025). Multilayer magnetic skyrmion devices for spiking neural networks. Neuromorphic Computing and Engineering, 5(1), 014005.

12. Jin, C., Song, C., Wang, J., & Liu, Q. (2016). Dynamics of antiferromagnetic skyrmion driven by the spin Hall effect. Applied Physics Letters, 109(18).

13. Zeissler, K., Finizio, S., Barton, C., Huxtable, A. J., Massey, J., Raabe, J., ... & Marrows, C. H. (2020). Diameter-independent skyrmion Hall angle observed in chiral magnetic multilayers. Nature communications, 11(1), 428.

14. Kläui, M. (2016). Skyrmion Hall effect revealed by direct time-resolved X-ray microscopy.

15. Yang, S., Moon, K. W., Ju, T. S., Son, J. W., Kim, T., Jeong, Y. J., ... & Hwang, C. (2025). Magnetic Skyrmion Neurons with Homeostasis for Spiking Neural Networks. ACS nano.

16. Torrejon, J., Riou, M., Araujo, F. A., Tsunegi, S., Khalsa, G., Querlioz, D., ... & Grollier, J. (2017). Neuromorphic computing with nanoscale spintronic oscillators. Nature, 547(7664), 428-431.

17. Zahedinejad, M., Awad, A. A., Muralidhar, S., Khymyn, R., Fulara, H., Mazraati, H., ... & Åkerman, J. (2020). Two-dimensional mutually synchronized spin Hall nano-oscillator arrays for neuromorphic computing. Nature nanotechnology, 15(1), 47-52.

18. Yang, S., Shin, J., Kim, T., Moon, K. W., Kim, J., Jang, G., ... & Hong, J. P. (2021). Integrated neuromorphic computing networks by artificial spin synapses and spin neurons. NPG Asia Materials, 13(1), 11.

19. Zhou, J., Zhao, T., Shu, X., Liu, L., Lin, W., Chen, S., ... & Chen, J. (2021). Spin–Orbit Torque-Induced Domain Nucleation for Neuromorphic Computing. Advanced Materials, 33(36), 2103672.

20. Wang, J. P., Shan, Z. S., Piramanayagam, S. N., & Chong, T. C. (2001). Anti-ferromagnetic coupling effects on energy barrier and reversal properties of recording media. IEEE transactions on magnetics, 37(4), 1445-1448.

21. Tang, K., Margulies, D., Polcyn, A., Supper, N., Do, H., Mirzamaani, M., ... & Xiao, Q. F. (2005). Laminated antiferromagnetically coupled media-optimization and extendibility. IEEE transactions on magnetics, 41(2), 642-647.

22. Piramanayagam, S. N., Wang, J. P., Hee, C. H., Pang, S. I., Chong, T. C., Shan, Z. S., & Huang, L. (2001). Noise reduction mechanisms in laminated antiferromagnetically coupled recording media. Applied Physics Letters, 79(15), 2423-2425.

23. Hee, C. H., Wang, J. P., Piramanayagam, S. N., & Chong, T. C. (2001). Thermal energy consideration in micromagnetic simulation for laminated antiferromagnetically coupled recording media. Applied Physics Letters, 79(11), 1646-1648.

24. Sekh, B., Kumar, D., Rahaman, H., Maddu, R., Chan, J., Mah, W. L. W., & Piramanayagam, S. N. (2024). Leaky-Integrate-Fire Neuron via Synthetic Antiferromagnetic Coupling and Spin-Orbit Torque. arXiv preprint arXiv:2408.08525.

25. Wang, D., Tang, R., Lin, H., Liu, L., Xu, N., Sun, Y., ... & Xing, G. (2023). Spintronic leaky-integrate-fire spiking neurons with self-reset and winner-takes-all for neuromorphic computing. Nature Communications, 14(1), 1068.

26. Liu, L., Wang, D., Wang, D., Sun, Y., Lin, H., Gong, X., ... & Liu, M. (2024). Domain wall magnetic tunnel junction-based artificial synapses and neurons for all-spin neuromorphic hardware. Nature Communications, 15(1), 4534.

27. Wang, Z. Q., Li, Z. Q., Sun, L., Zhang, Z. Y., He, K., Niu, H., ... & Miao, B. F. (2024). Inverse spin Hall effect dominated spin-charge conversion in (101) and (110)-oriented RuO 2 films. Physical Review Letters, 133(4), 046701.

28. Li, Z., Zhang, Z., Chen, Y., Hu, S., Ji, Y., Yan, Y., ... & Lu, X. (2025). Fully Field-Free Spin-Orbit Torque Switching Induced by Spin Splitting Effect in Altermagnetic RuO2. Advanced Materials, 37(12), 2416712.

29. Karube, S., Tanaka, T., Sugawara, D., Kadoguchi, N., Kohda, M., & Nitta, J. (2021). Observation of spin-splitter torque in collinear antiferromagnetic RuO$_2$. arXiv preprint arXiv:2111.07487.

30. Šmejkal, L., González-Hernández, R., Jungwirth, T., & Sinova, J. (2020). Crystal time-reversal symmetry breaking and spontaneous Hall effect in collinear antiferromagnets. Science advances, 6(23), eaaz8809.

31. Sinova, J., Valenzuela, S. O., Wunderlich, J., Back, C. H., & Jungwirth, T. (2015). Spin hall effects. Reviews of modern physics, 87(4), 1213-1260.

32. Bose, A., Schreiber, N. J., Jain, R., Shao, D. F., Nair, H. P., Sun, J., ... & Ralph, D. C. (2022). Tilted spin current generated by the collinear antiferromagnet ruthenium dioxide. Nature Electronics, 5(5), 267-274.

33. Sekh, B., Rahaman, H., Maddu, R., Mishra, P. K., Jin, T., & Piramanayagam, S. N. (2025). Enhanced Field-Free Perpendicular Magnetization Switching via spin splitting torque in Altermagnetic RuO2-based Heterostructures. arXiv preprint arXiv:2501.12593.

34. Karube, S.; Tanaka, T.; Sugawara, D.; Kadoguchi, N.; Kohda, M.; Nitta, J. Observation of Spin-Splitter Torque in Collinear Antiferromagnetic $RuO_2$. Phys. Rev. Lett. 2022, 129, 137201.

35. Vansteenkiste, A. *et al*. The design and verification of MuMax3. *AIP Adv.* **4**, 107133 (2014).

36. Sekh, B. *et al*. Enhanced Field-Free Perpendicular Magnetization Switching via spin splitting torque in Altermagnetic RuO2-based Heterostructures. Preprint at https://doi.org/10.48550/arXiv.2501.12593 (2025).

37. Song, Y. *et al*. Field-Free Spin-Orbit Torque Switching of Perpendicular Magnetization by Making Full Use of Spin Hall Effect. *Adv. Electron. Mater.* **9**, 2200987 (2023).

38. Koyama, T. *et al*. Observation of the intrinsic pinning of a magnetic domain wall in a ferromagnetic nanowire. *Nat. Mater.* **10**, 194–197 (2011).

39. Christensen, D. V. *et al*. 2022 roadmap on neuromorphic computing and engineering. *Neuromorphic Comput. Eng.* **2**, 022501 (2022).

40. Orchard, G., Jayawant, A., Cohen, G. K. & Thakor, N. Converting Static Image Datasets to Spiking Neuromorphic Datasets Using Saccades. Front. Neurosci. 9, (2015).

41. Maes, A., Barahona, M. & Clopath, C. Learning spatiotemporal signals using a recurrent spiking network that discretizes time. *PLoS Comput. Biol.* **16**, e1007606 (2020).

42. Garrido, J. A., Ros, E. & D'Angelo, E. Spike Timing Regulation on the Millisecond Scale by Distributed Synaptic Plasticity at the Cerebellum Input Stage: A Simulation Study. *Front. Comput. Neurosci.* **7**, (2013).

43. Bouvier, M. *et al.* Spiking Neural Networks Hardware Implementations and Challenges: A Survey. *J Emerg Technol Comput Syst* **15**, 22:1-22:35 (2019).

44. Iyer, L. R. & Basu, A. Unsupervised learning of event-based image recordings using spike-timing-dependent plasticity. in *2017 International Joint Conference on Neural Networks (IJCNN)* 1840–1846 (2017). doi:10.1109/IJCNN.2017.7966074.

45. Iyer, L. R., Chua, Y. & Li, H. Is Neuromorphic MNIST Neuromorphic? Analyzing the Discriminative Power of Neuromorphic Datasets in the Time Domain. *Front. Neurosci.* **15**, (2021).